\documentclass{article}

\usepackage{arxiv}
\usepackage{float}
\usepackage{multirow}
\usepackage[utf8]{inputenc} 
\usepackage[T1]{fontenc}    
\usepackage{hyperref}       
\usepackage{url}            
\usepackage{booktabs}       
\usepackage{amsfonts}       
\usepackage{nicefrac}       
\usepackage{microtype}      
\usepackage{lipsum}
\usepackage{amsmath} 
\usepackage{graphicx}
\graphicspath{ {./images/} }
\usepackage{array}

\title{Interpretable Gallbladder Ultrasound Diagnosis: A Lightweight Web–Mobile Software Platform with Real-Time XAI}

\author{
Fuyad Hasan Bhoyan \\
  Department of Computer Science and Engineering\\ University of
 Liberal Arts Bangladesh\\
  Dhaka, Bangladesh \\
  \texttt{fuyad.hasan.cse@ulab.edu.bd} \\
  \And
Prashanta Sarker\\
Department of Computer Science and Engineering\\ University of
 Liberal Arts Bangladesh\\
  Dhaka, Bangladesh \\
  \texttt{prashanta.sarker.cse@ulab.edu.bd} \\
  \And
Parsia Noor Ethila\\
Department of Computer Science and Engineering\\ University of
 Liberal Arts Bangladesh\\
  Dhaka, Bangladesh \\
  \texttt{parsia.noor.cse@ulab.edu.bd} \\
    \And
Md. Emon Hossain\\
Department of Computer Science and Engineering\\ University of
 Liberal Arts Bangladesh\\
  Dhaka, Bangladesh \\
  \texttt{emon.hossain.cse@ulab.edu.bd} \\
       \And
Md Kaviul Hossain \\
  Department of Computer Science and Engineering\\ University of
 Liberal Arts Bangladesh\\
  Dhaka, Bangladesh \\
  \texttt{kaviul.hossain@ulab.edu.bd} \\
     \And
 Md Humaion Kabir Mehedi \\
  Department of Computer Science and Engineering\\ BRAC University\\
  Dhaka, Bangladesh  \\
  \texttt{humaion.kabir.mehedi@g.bracu.ac.bd} \\
}

\begin{document}
\maketitle
\begin{abstract}
Early and accurate detection of gallbladder diseases is crucial, yet ultrasound interpretation is challenging. To address this, an AI-driven diagnostic software integrates our hybrid deep learning model MobResTaNet to classify ten categories, nine gallbladder disease types and normal directly from ultrasound images. The system delivers interpretable, real-time predictions via Explainable AI (XAI) visualizations, supporting transparent clinical decision-making. It achieves up to 99.85\% accuracy with only 2.24M parameters. Deployed as web and mobile applications using HTML, CSS, JavaScript, Bootstrap, and Flutter, the software provides efficient, accessible, and trustworthy diagnostic support at the point of care.
\end{abstract}


\section{Introduction}
\begin{figure}[htbp]
\includegraphics[width=0.9\textwidth]{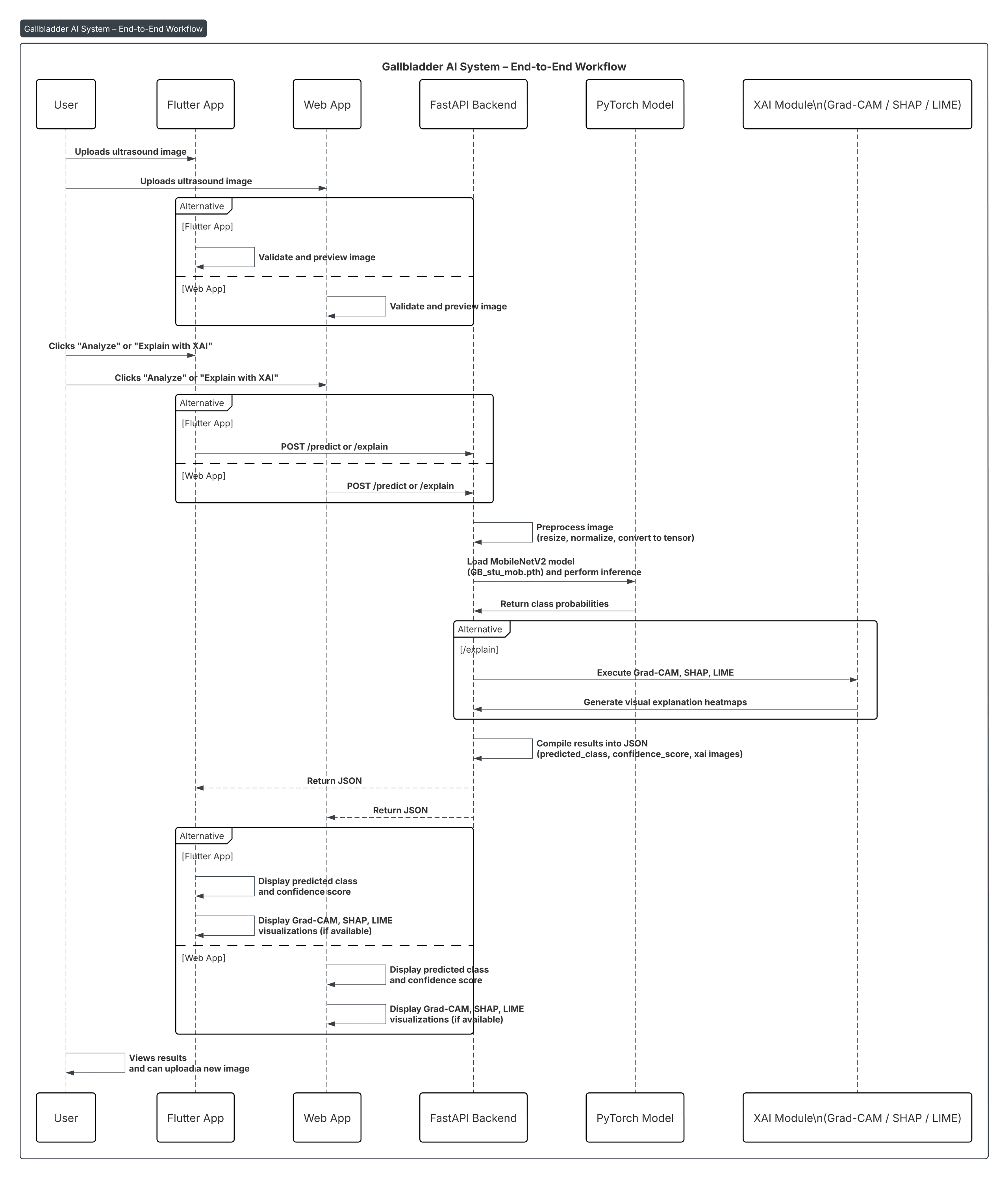}
    \caption{Sequence Diagram of the software workflow}
    \label{fig:d1}
\end{figure}

 Gallbladder (GB) diseases, including gallstones, cholecystitis, adenomyomatosis, carcinoma, wall thickening, and polyps, are significant causes of abdominal pain and biliary dysfunction \cite{1}. If not detected early, these conditions can result in severe complications such as infection and cancer. According to GLOBOCAN 2020, gallbladder cancer accounts for 115,949 global cases and 84,695 annual deaths, ranking as the seventh most common digestive tract tumor \cite{2}. Studies have indicated that 3-8\% of GB polyps are malignant \cite{3}, while up to 30\% of acute cholecystitis cases progress to gangrenous cholecystitis \cite{4}. Despite the widespread use of ultrasound (UI), computed tomography (CT), and magnetic resonance imaging (MRI) for diagnosis, their interpretation heavily relies on radiologists' expertise, often leading to variability in results. Recent advancements in artificial intelligence (AI) and deep learning (DL) have demonstrated strong potential for automating and enhancing medical image interpretation \cite{5}. A notable example is COV-ADSX, an automated web-based diagnostic system that integrates deep learning and XGBoost within the Django framework for COVID-19 detection  \cite{6}. The system enables users to upload X-ray images, extract image features using a pre-trained CNN, perform classification with XGBoost, and apply Grad-CAM visualization to display decision regions, allowing clinicians to verify the model’s focus on the correct regions. Inspired by such an intelligent diagnostic system, this study introduced a deep-learning-based software for gallbladder disease diagnosis. Developed with a hybrid CNN model named ‘MobResTaNet’ and using HTML, CSS, JavaScript, Bootstrap, and Flutter, it supports Explainable AI (XAI) visualizations to deliver interpretable, real-time, and reliable diagnostic results for clinical use \cite{7}.
 Figure \ref{fig:d1} illustrates the end-to-end workflow of the proposed web–mobile gallbladder ultrasound diagnostic system, highlighting the interactions between the user interface, backend services, deep learning model, and explainability modules.
\section{High-Level Description}
This application leverages AI to assist in the analysis of gallbladder ultrasound images. It allows clinicians, trainees, and researchers to either upload or capture an image, providing an instant prediction across ten categories, which include nine types of gallbladder diseases and a normal category. In addition, it offers explainable visualizations that clarify the decision-making process of the model. The aim is to alleviate the burden of interpretation, standardize the triage process, and deliver transparent and reproducible support at the point-of-care.

\section{Data and Evaluation}
The model incorporated in this software was trained and evaluated using two publicly available gallbladder ultrasound datasets: UIdataGB and GBCU \cite{8},\cite{9}. The combined dataset encompassed ten classes, comprising nine gallbladder disease categories and one normal category. UIdataGB was sourced from four medical institutions in Baghdad, Iraq: Medicine Teaching Hospital, Al-Numan Teaching Hospital, Specialized Gastroenterology Center, and Jenin Hospital, employing Siemens Acuson X700, Philips Affiniti 70, Philips CX50, and Canon Viamo c100 ultrasound machines, respectively. Expert radiologists annotated and verified the data according to the diagnostic criteria. The GBCU dataset, obtained at PGIMER, Chandigarh, India, includes biopsy-confirmed labels for normal, benign, and malignant categories using the Logiq S8 ultrasound machine. Both datasets received ethical approval and adhered to de-identification and patient consent protocols. The images were resized to 224 × 224 pixels, converted to tensors, and normalized to ensure stable training. Data augmentation involved random horizontal and vertical flips (p=0.3-0.7), rotations (15°–60°), and affine transformations to enhance generalization. Each class comprised 2000 samples, divided into 1200 training, 400 validation, and 400 testing images, following a 60:20:20 split to ensure a balanced representation and patient-level separation. This preprocessing facilitated standardized and diversified input data, thereby supporting a robust and reproducible model performance evaluation.

\section{Software Overview}

The developed software system is a clinician-oriented diagnostic application designed to facilitate the interpretation of gallbladder ultrasound images using artificial intelligence. The platform employs the proposed MobResTaNet model to classify input images into ten categories, encompassing nine distinct gallbladder abnormalities and one normal condition, with real-time inferences and explainable outputs. Users, including radiologists and medical practitioners, can upload or capture ultrasound frames directly using a web-based interface or mobile application. Upon processing, the system delivers immediate classification results accompanied by Explainable AI (XAI) visualizations, such as Grad-CAM, SHAP, and LIME overlays, which highlight the specific image regions influencing the model’s decision. This feature enhances clinical trust and supports the informed validation of AI predictions. The application prioritizes accessibility, featuring a simple, responsive user interface and reproducibility through Dockerized deployment, facilitating seamless integration into various healthcare settings. Its transparency and interpretability render it particularly suitable for decision support in radiological diagnostics, training purposes for medical students, and clinical research environments exploring AI-assisted imaging applications.
\subsection{Functionality \& Workflow}
\subsubsection{Core capabilities}
The system provides:
\begin{itemize}
    \item Image ingestion from web or mobile.

    \item Automated classification using the hybrid deep learning model.
    \item Real-time XAI with Grad-CAM, SHAP, and LIME overlays.
    \item Confidence scores, and one-click export for reports or teaching materials.
\end{itemize}

\subsubsection{Typical workflow}
\begin{itemize}
    \item Users load an ultrasound frame.
    \item The backend performs preprocessing and inference.

    \item The UI displays the predicted class, confidence, and XAI overlays.
    \item Users can adjust overlay opacity, compare methods, and export results.
\end{itemize}

\subsubsection{Platforms and UI}
\begin{itemize}
    \item \textbf{Web:} HTML, CSS, JavaScript, Bootstrap UI for a clean, device-agnostic interface (desktop/tablet).
    \item \textbf{Mobile:} Flutter app for bedside use, supports remote inference via API and local caching of results when connectivity is limited.

    \item \textbf{Back-end and Deployment:} The inference service was implemented using FastAPI (Python) in conjunction with PyTorch for model execution. A Docker image encapsulates the runtime environment (CPU or GPU), facilitating reproducible deployment across hospital servers, research laboratories and cloud instances. Health checks and logging endpoints were incorporated to support operations at scale. 
    \item \textbf{Explainability Engine:} Integrated Grad-CAM, SHAP (GradientExplainer), and LIME (superpixel) visualizations offer complementary insights into model reasoning. These overlays are consistently color-mapped and can be exported as PNG/base64 files for inclusion in electronic health record notes, presentation slides, or audit trails.

    \item \textbf{Scope and Performance:} The system supported 10-class inference (nine diseases plus normal). Utilizing the integrated MobResTaNet model (2.24M parameters), achieved an accuracy of up to 99.85\% in internal evaluations while maintaining low latency suitable for interactive use.
\end{itemize}

\section{Implementation Details}
\begin{itemize}
    \item \textbf{Model integration:} The application incorporates the MobResTaNet architecture and its student version using Torch checkpoints. Before inference, the images underwent preprocessing, which included standardized resizing to 224×224, normalization, and batch collation. A single prediction interface provides class probabilities and top-1 labels for explainability modules.
    \item\textbf{Explainability pipeline:}  Grad-CAM visualizes class-discriminative regions by extracting the activation gradients from the last convolutional layer. SHAP uses PartitionExplainer with an inpainting-based image masker ("inpaint\_telea") to achieve stable, detailed pixel attribution in relation to the model’s background expectations. LIME iteratively perturbs superpixels to assess feature importance and create interpretable regional heatmaps. All the explainers utilized a unified overlay renderer with Turbo colormaps and adaptive transparency to maintain a balanced contrast while preserving the brightness of the original image.

    \item \textbf{Systems engineering:} The system is organized into modular services, including model inference, XAI routines, and API endpoints, all of which are managed within a FastAPI application. The CORS middleware, structured JSON logging, and /health endpoint offer monitoring and integration options for cloud deployments. Environment-driven configuration sets runtime parameters (device selection, overlay intensity, and thresholds) to ensure consistency between the development and production containers.  
    \item \textbf{Security \& privacy:} Incoming requests are rigorously validated to accept only image MIME types. All processing occurs in memory, avoiding persistent disk writes. Optional runtime flags can enforce temporary file cleanup and limit I/O for sensitive environments. Deployments can implement container-level role-based network policies or integrate site-specific governance frameworks for research or clinical compliance.
\end{itemize}

\section{Impact Overview}
The software facilitates large-scale, class-stratified attribution analysis across ten categories, allowing for the examination of dataset shifts, scanner variability, and labeling quality issues. It standardizes case reviews with side-by-side XAI, accelerates reader studies, and provides reproducible exports for protocol discussions and curriculum development. Clinicians can quickly obtain a clear, interpretable second opinion, while trainees can enhance their learning by linking visual attributions with anatomical knowledge. The platform is designed for radiology and hepatobiliary clinics, teaching hospitals, and academic laboratories. Docker packaging and a straightforward REST interface simplify the integration into existing workflows and research pipelines. The service can be incorporated into PACS/RIS viewers using lightweight adapters or licensable SDKs, with both on-premises and managed cloud models available.

\section{Ongoing Work Using the Software}

The model currently integrated into the software is trained solely using a supervised method, which heavily depends on labeled data. The authors are working on implementing a semi-supervised approach to enhance generalization and reduce the burden of data annotation.

\section{Limitations and Future Improvements}

\begin{itemize}
    \item \textbf{Generalization:} Performance may degrade under domain shifts (e.g., new scanners, protocols, and populations) or poor-quality inputs.
    \item\textbf{Class imbalance:}  Real-world prevalence may differ from training distributions, and probability calibration should be monitored.

    \item \textbf{Explainability caveats:} Saliency reflects associations, not clinical causation, and different XAI methods can yield divergent results. In the future, approaches such as the dynamic explainability module selector will be implemented.  
    \item \textbf{Regulatory status:} The software is not a certified medical device; its clinical use requires appropriate oversight, validation, and approval.
    \item \textbf{Data and formats:} DICOM ingestion, cine-loop support, and anonymization utilities.

    \item \textbf{Uncertainty and calibration:} Per-case uncertainty scores and temperature scaling for more reliable thresholds.

    \item \textbf{Learning schemes:} Continual/federated learning and semi-supervised labeling assistance to reduce the annotation burden.

    \item \textbf{Workflow integration:} PACS/RIS connectors, template report generation, and audit dashboards to support service-level monitoring.

\end{itemize}
\section{Conclusion}
Our Web-Mobile software embedded with the hybrid model- MobResTaNet, provides efficient and robust disease interpretation of GallBladder ultrasound images with a whopping accuracy of 99.85\%. With only 2.24M parameters, it displays promising results, which is not only astounding but also provides ,solid ground for fast and effective gall bladder disease detection. It is an innovation that holds the potential to bring an unforeseen revolution to the medical field. Developed with a state-of-the-art tech stack, our software leverages large-scale class-stratified attribution analysis across nine categories of gall bladder diseases. Although it is not yet certified medical equipment, and its application requires appropriate oversight and supervision, with regular monitoring and medical validation, our web app has the potential to revolutionize Gallbladder disease diagnosis based on Ultrasound images.

\section*{Code metadata}
\textit{Please replace the italicized text in the right column with the correct information about your code/software and leave the left column untouched.}\\

\noindent
\begin{tabular}{|l|p{6.5cm}|p{9.5cm}|}
\hline
\textbf{Nr.} & \textbf{Code metadata description} & \textbf{Please fill in this column} \\
\hline
C1 & Current code version & \textit{v1.1.1} \\
\hline
C2 & Permanent link to code/repository used for this code version & \url{https://github.com/Prashanta4/gallbladder-web} \\
\hline
C3  & Permanent link to Reproducible Capsule & \\
\hline
C4 & Legal Code License   & MIT License. \\
\hline
C5 & Code versioning system used & \textit{GitHub} \\
\hline
C6 & Software code languages, tools, and services used & \textit{Python 3.6 or later, HTML, CSS, JavaScript, Docker, Fast API}\\
\hline
C7 & Compilation requirements, operating environments \& dependencies & Not available, because all the requirements are fulfilled in the Docker image.\\
\hline
C8 & If available Link to developer documentation/manual & \url{https://github.com/Prashanta4/gallbladder-web/blob/main/README.md}\\
\hline
C9 & Support email for questions & fuyadhasanbhoyan@gmail.com\\
\hline
\end{tabular}\\
\vskip0.5cm
\noindent


 \section*{Funding}
No funding was received for conducting this research or for preparing this manuscript.

\section*{Declaration of Competing Interest}
The authors declare that they have no known competing financial interests or personal relationships that could influence the work reported in this study.


\end{document}